\documentclass[aps,prl,twocolumn,amsmath,amssymb,superscriptaddress]{revtex4}

\usepackage{amssymb}
\usepackage{amsmath}
\usepackage{graphicx}
\usepackage{subfigure}
\usepackage{textcomp}
\usepackage{color}
\usepackage{amsfonts}
\usepackage{bbold}
\usepackage{dsfont}
\usepackage{epsfig}
\usepackage{hyperref}

\begin{document}

\title{The ground state of a spin-crossover molecule calculated by diffusion Monte Carlo}

\author{A. Droghetti}
\affiliation{School of Physics and CRANN, Trinity College, Dublin 2, Ireland}
\author{D. Alf\`e}
\affiliation{London Centre for Nanotechnology, Dept. of Earth Sciences, 
Deptartment of Physics and Astronomy,
University College London,
Gower Street, London, WC1E 6BT,  UK}
\author{S. Sanvito}
\affiliation{School of Physics and CRANN, Trinity College, Dublin 2, Ireland}

\date{\today}

\begin{abstract}
Spin crossover molecules have recently emerged as a family of compounds potentially useful for implementing molecular spintronics 
devices. The calculations of the electronic properties of such molecules is a formidable theoretical challenge as one has to describe
the spin ground state of a transition metal as the legand field changes. The problem is dominated by the interplay between strong 
electron correlation at the transition metal site and charge delocalization over the ligands, and thus it fits into a class of problems 
where density functional theory may be inadequate. Furthermore, the crossover activity is extremely sensitive to environmental 
conditions, which are difficult to fully characterize. Here we discuss the phase transition of a prototypical spin crossover molecule 
as obtained with diffusion Monte Carlo simulations. We demonstrate that the ground state changes depending on whether the 
molecule is in the gas or in the solid phase. As our calculation provides a solid benchmark for the theory we then assess the 
performances of density functional theory. We find that the low spin state is always over-stabilized, not only by the (semi-)local functionals, 
but even by the most commonly used hybrids (such as B3LYP and PBE0). We then propose that reliable results can be obtained by using 
hybrid functionals containing about $50\%$ of exact-exchange. 
\end{abstract}

\maketitle

In nature there is a vast class of molecules whose magnetic moment can be altered by an external stimulus. Typical examples of such molecules 
are the spin-crossover (SC) complexes \cite{Kahn,SC_molecules}, which, in their most abundant form, contain a Fe$^{2+}$ ion in octahedral 
coordination~\cite{note} and exhibit a transition from the low spin (LS) (singlet) ground state to a high spin (HS) (quintet) metastable state. 
Other examples are the cobalt dioxolene molecules \cite{Sato,Hendrickson,Dei}. These undergo the so-called valence tautomeric interconversion 
(VTI), namely an interconversion between two redox isomers, which differ in charge distribution and spin configuration. Both the SC transition and 
the VTI are usually observed for molecules in a single crystal and can be triggered by variations in temperature and pressure or by optical 
irradiation \cite{Hauser}. Furthermore it was also recently suggested that the VTI \cite{me} and the spin ground state of a two-center polar 
molecule \cite{NB} can be controlled by a static electric field . 

SC complexes are promising materials candidates for molecular spintronics applications \cite{bogani,stefano_mol_spin}. Devices incorporating 
such molecules are predicted to display drastic changes in the current-voltage curve across the phase transition \cite{Nadjib,Aravena}, 
and several transport experiments have recently achieved encouraging results. Alam {\it et al.} \cite{Alam} were able to distinguish the 
spin state of a SC molecule placed on graphite by scanning tunnel microscopy, while Prins {\it et al.} \cite{Prins} demonstrated that 
the temperature dependent conductance of a device incorporating a SC cluster correlates well with the phase transition. In other cases, 
however, the data are not easy to interpret \cite{Ruben} and the experimental investigations are combined with density functional theory 
(DFT) simulations. In principle DFT should allow the computation of quantities not easily accessible by experiments and should also provide 
parameters for effective transport models. However, unfortunately DFT results for SC molecules depend strongly (even qualitatively) on 
the choice of the exchange-correlation functional used \cite{Swart,Zein,Fouqueau_1,Fouqueau_2,Pierloot1,Pierloot2} and no standard 
has yet emerged. This essentially means that DFT is still not a predictive theory for this problem.

Since most of the local and semilocal functionals underestimate the exchange energy, they tend to favor the LS state against the HS 
one~\cite{Swart,Zein}. This shortcome often leads to such large errors that even stable HS molecules are described as LS~\cite{Fouqueau_1,Fouqueau_2}. 
In contrast, the most commonly used hybrid functionals are believed to over-stabilize the HS state~\cite{Reiher_1, Reiher_2}. Besides, several 
authors~\cite{Zein,Fouqueau_1,Fouqueau_2, me2} have criticized the common practice, that consists in assessing the performances of the 
various functionals by direct comparison with the experimental data. In fact, while experiments are usually performed for molecules in the 
condensed phase, DFT results refer to molecules in the gas phase. Since the properties of SC complexes depend strongly on environmental 
conditions (counter-ions, interaction between molecules, ``strain effects'' etc...) \cite{Robert_1,Robert_2} their ground state may not 
be the same in these two phases. The question then becomes: {\it can we produce a robust benchmark for DFT against the problem of predicting
the Physics of SC compounds?} In order to answer to this question {\it ab-initio} methods more accurate than DFT have to be considered.

In the past wave-function based methods were used for this problem \cite{Fouqueau_1,Fouqueau_2,Pierloot1,Pierloot2,Ruben}.
However, as the authors themselves pointed out, the results were plagued by systematic errors ascribed to the limited basis set used for 
Fe$^{2+}$ and by the fact that the methods themselves neglect dynamic correlation (although this can be partly accounted for through a perturbative treatment). Here we propose an alternative route and
perform diffusion Monte Carlo (DMC) \cite{Foulkes} calculations for a prototypical Fe$^{2+}$ spin crossover molecule. As DMC 
represents one of the most accurate electronic structure method currently available in order to compute ground state energies, our calculations provided a solid benchmark for 
assessing the performances of DFT.
\begin{figure}[ht]\centering
\centering \includegraphics[scale=0.3,clip=true]{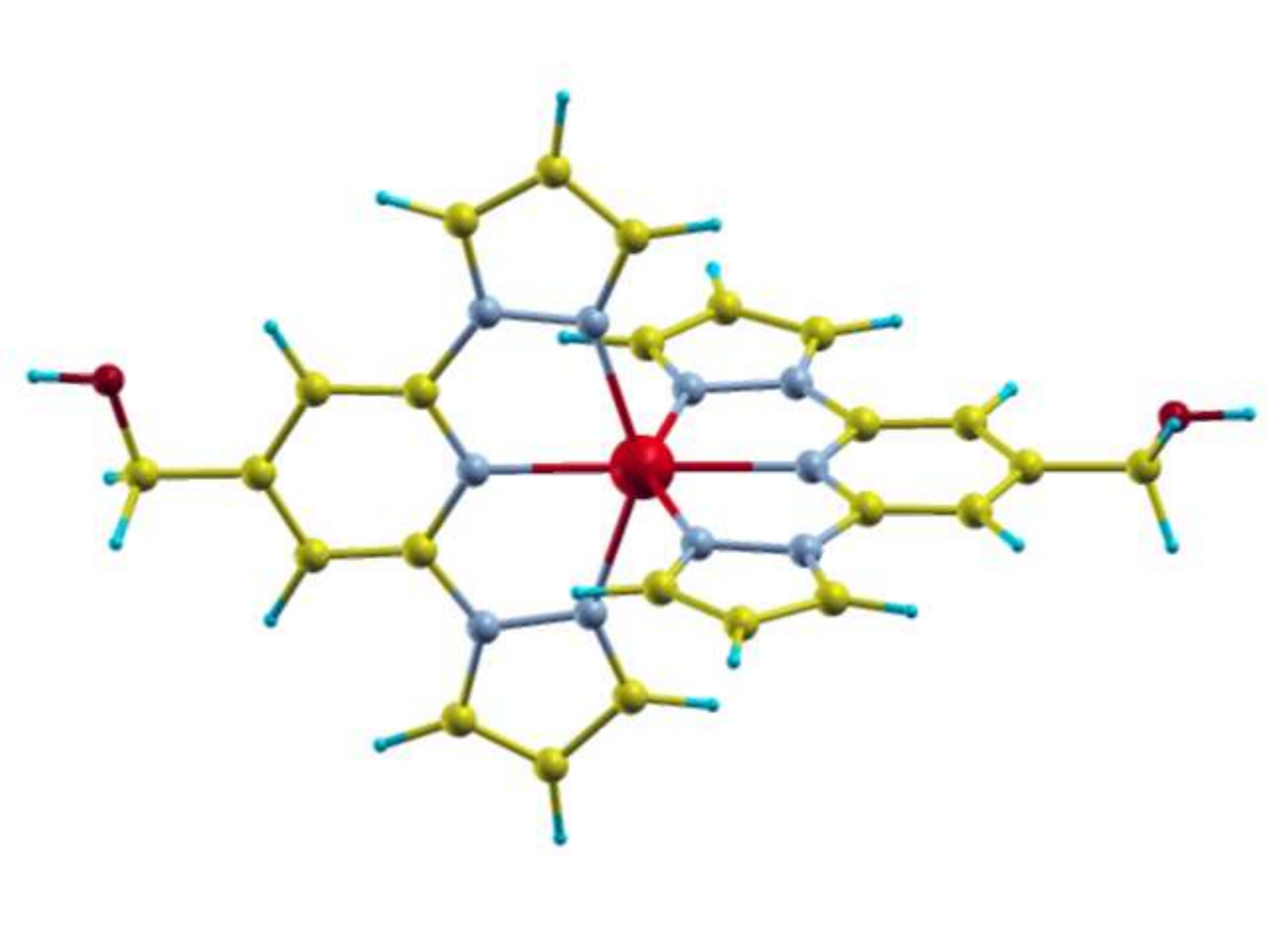}
\caption{(Color on line) The cationic unit [FeL$_2$]$^{2+}$ (L=2,6-dypirazol-1-yl-4-hydroxymethylpyridine) used in the DMC calculations. 
Color code: C=yellow, O=red (small sphere), Fe=red (large sphere), N=grey, H=blue.}\label{pic_molecule}
\end{figure}
In particular we consider the molecule [FeL$_2$](BF$_4$)$_2$ (L=2,6-dypirazol-1-yl-4-hydroxymethylpyridine) \cite{Money} (see Fig.~\ref{pic_molecule}).
We show that its ground state in the gas phase is HS but that a phase transition may exist in the solid state due to a number of crystal-related effects.
We then show that the same result can be obtained by DFT hybrid functionals containing approximately $50\%$ of exact exchange, thus confirming
early calculations for model molecules~\cite{me2}. This establishes a recipe for the use of DFT for this class of materials, and it opens the opportunity
to investigate with confidence the spin crossover transition of molecules in different environments (for instance on surfaces).

DFT calculations are performed with the {\sc nwchem} code \cite{nwchem}. We use several functionals belonging to different classes: 
1) the Vosko-Wilk-Nussair local density approximation (LDA) \cite{Vosko}; 2) the generalized gradient approximation BP86, which combines 
the Becke88 exchange functional \cite{Becke} with the Predew86 correlation one \cite{Perdew86}; 3) the hybrid functionals B3LYP \cite{b3lyp}, 
PBE0 \cite{Ernzerhof,Adamo} and the Becke-HH \cite{becke_HH}, which include respectively $20\%$, $25\%$ and $50\%$ of exact exchange. We also consider a re-parametrization of the B3LYP functional, called B3LYP$^*$, which includes 
only $15\%$ of HF exchange. This was introduced by Reiher and co-workes specifically in order to describe Fe$^{2+}$ complexes \cite{Reiher_1,Reiher_2}.
The Ahlrichs triple-zeta polarized basis set \cite{Ahlrichs} is used throughout. DMC calculations are performed by using the {\sc casino} 
code \cite{casino}. The imaginary time evolution of the Schr\"odinger equation has been performed with the usual short time approximation and 
time-steps of $0.0125$ and $0.005$ a.u. are used. Dirac-Fock pseudopotentials \cite{Trail1,Trail2} with the ``potential localization 
approximation'' \cite{Mitas} have been used. The single-particle orbitals of the trial wave function are obtained through (LDA) DFT calculations 
performed with the plane-wave (PW) code {\sc quantum espresso} \cite{espresso}. The same pseudopotentials used for the DMC calculations 
are employed. The PW cutoff is fixed at $300$ Ry and the PW are re-expanded in terms of B-splines \cite{Dario}. The B-spline grid spacing is 
$a = \pi/ G_{\mathrm{max}}$, where $G_{\mathrm{max}}$ is the length of the largest vector employed in the PW calculations. Periodic boundary 
conditions are employed for the PW-DFT calculations and supercells as large as $40$~\AA~ are considered. In contrast, no periodic boundary 
conditions are imposed with DMC. Counter ions have been ignored and calculations are presented for the cation (Fig. \ref{pic_molecule}) to 
which we will  generally refer as the molecule.

\begin{figure}[ht!]\centering
\centering \includegraphics[scale=0.25,clip=true]{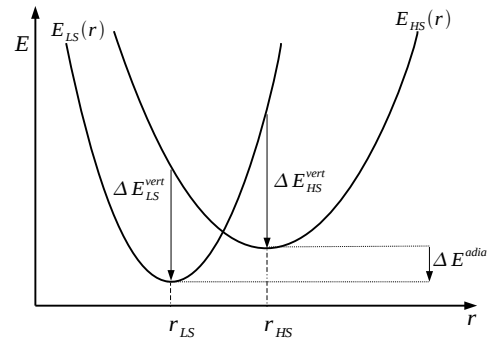}
\caption{Potential energy surface of the HS and LS state of a SC molecule. The collective coordinate $r$ represents all of the nuclear 
coordinates of the molecule. The adiabatic energy gap, $\Delta E^{\mathrm{adia}}$, and the vertical energy gaps, 
$\Delta E^{\mathrm{vert}}_\mathrm{LS}=\Delta E^{\mathrm{vert}}(r_\mathrm{LS})$ and 
$\Delta E^{\mathrm{vert}}_\mathrm{HS}=\Delta E^{\mathrm{vert}}(r_\mathrm{HS})$ are also indicated.}\label{adia_pic}
\end{figure}
The crucial quantity for understanding the spin crossover transition is the potential energy surface, schematically displayed in Fig.~\ref{adia_pic}. This is 
typically plotted for the two different spin configurations as a function of a collective reaction coordinate $r$, which interpolates the molecule geometry 
along the LS to HS phase transition. In our case DMC and DFT are used to compute the ``adiabatic energy gap'' \cite{Zein} defined as
\begin{equation}
\Delta E^{\mathrm{adia}}=E_\mathrm{HS}(r_\mathrm{HS})-E_\mathrm{LS}(r_\mathrm{LS})\,,
\end{equation}
where $r_\mathrm{LS}$ ($r_\mathrm{HS}$) and $E_\mathrm{LS}(r_\mathrm{LS})$ [$E_\mathrm{HS}(r_\mathrm{HS})$] represent respectively 
the geometry and the total energy of the LS-singlet (HS-quintet) state. When studying SC molecules at zero temperature $\Delta E^{\mathrm{adia}}$ 
is the central quantity, as it indicates whether the molecule ground state is LS ($\Delta E^{\mathrm{adia}}>0$) or HS ($\Delta E^{\mathrm{adia}}<0$). 
We also calculate the ``vertical energy gaps''\cite{Zein} 
\begin{eqnarray}
&\Delta E^{\mathrm{vert}}(r_\mathrm{LS})=E_\mathrm{HS}(r_\mathrm{LS})-E_\mathrm{LS}(r_\mathrm{LS})\label{delta_E_vert1}\,,\\
&\Delta E^{\mathrm{vert}}(r_\mathrm{HS})=E_\mathrm{HS}(r_\mathrm{HS})-E_\mathrm{LS}(r_\mathrm{HS})\label{delta_E_vert2}\,,
\end{eqnarray}
where $E_\mathrm{HS}(r_\mathrm{LS})$ [$E_\mathrm{LS}(r_\mathrm{HS})$] is the energy of the quintet (singlet) state for the $r_{LS}$ ($r_{HS}$)
geometry (see Fig. \ref{adia_pic}).

\begin{table}[h!]\centering
\scalebox{0.85}{%
\begin{tabular}{lcc}\hline\hline
 & Spin state & $d$ (\AA) \\\hline\hline
BP86 &  LS & $1.898(2), 1.958(4)$  \\
BP86  & HS  & $2.118, 2.127, 2.159,2.166, 2.185(2)$\\
B3LYP* & LS & $1.923(2)$, $1.996(4)$\\
B3LYP* & HS & $2.182(2),2.211(2),2.207(2)$\\
B3LYP & LS& $1.933(2),2.011(4)$ \\
B3LYP & HS &  $2.185(2),2.218(2), 2.222(2)$\\
PBE0  &  LS &  $1.910 (2), 1.984(4)$\\
PBE0 & HS & $2.169(2),2.1965(4)$ \\
Exp. & LS  & $ 1.909,1.991,1.912,1.985,1.980,1.992$ \\
Exp. & HS  & $2.105,2.163, 2.103,2.160,2.170,2.153$ \\
\hline
\end{tabular}}
\caption{Experimental and calculated Fe-N bond-lengths for the [FeL$_2$]$^{2+}$ cation. The number of bonds of a given length are indicated inside the 
bracket. The average difference between HS and LS Fe-N bond-lengths is about $0.2$\AA, a typical value for SC molecules.}  \label{Tab_bond_SC_molecule}
\end{table}
DMC calculations were first carried out by using the molecular geometries optimized with DFT for the molecule in the gas phase. The Fe-ligand bond lengths 
computed with the various functionals are listed in Tab.~\ref{Tab_bond_SC_molecule}. As the DMC energy differences between the geometries calculated
from BP86, B3LYP*, B3LYP and PBE0 are of the same order of magnitude as the Monte Carlo statistical error, we have not been able to firmly establish which functional 
produces the best structure. We have then decided to present results only for the structures relaxed with B3LYP, keeping in mind that the same are essentially 
valid also for BP86 and PBE0.

\begin{table}[h!]\centering
\scalebox{0.8}{%
\begin{tabular}{lccc}\hline\hline
Method  & $\Delta E^{\mathrm{vert}}(r_\mathrm{LS})$ (eV) & $\Delta E^{\mathrm{vert}}(r_\mathrm{HS})$ (eV) &  $\Delta E^{\mathrm{adia}}$ (eV)\\ \hline\hline
BP86  & $2.87$ & $-0.180$ &$1.23$ \\
B3LYP* &$1.97$& $-1.013$ & $0.331$\\
B3LYP  & $1.54$ & $-1.23$& $0.012$\\
PBE0  & $1.19$ & $-1.74$ & $-0.23$\\
Becke-HH    & $0.51$ & $-2.50$  &  $-1.33$\\
DMC ($\Delta \tau=0.005$ a.u)& $0.28(4)$ & $-2.57(4)$ & $-1.19(4)$\\
\hline
\end{tabular}}
\caption{Adiabatic and vertical energy gaps for the [FeL$_2$]$^{2+}$ cation calculated with DFT and DMC at the DFT-B3LYP relaxed geometry.
The relative Monte Carlo statistical error is indicated in bracket.} 
\label{energy_SC_molecule_theory}
\end{table}
The DMC adiabatic energy gap is reported in Tab. \ref{energy_SC_molecule_theory}. Our result indicates that the molecule in the gas phase is in its
high-spin state, in contrast to the common belief and to the experimental result for the single crystal. Such ground state is indeed quite robust as DMC 
gives us an adiabatic energy gap of -1.20~eV. Since DMC provides a unequivocal assignment of the molecule ground state, it essentially establishes that
no spin crossover transition is expected for  [FeL$_2$]$^{2+}$ in the gas phase. Hence, in order to account for the experimentally observed SC transition, 
one needs to understand how the embedding of the molecule in a crystal is able to reverse the relative order of the HS and the LS states at zero-temperature,
i.e. to change the sign of $\Delta E^{\mathrm{adia}}$.

\begin{table}[h!]\centering
\scalebox{0.8}{%
\begin{tabular}{lccc}\hline\hline
  & $\Delta E^{\mathrm{vert}}(r_\mathrm{LS})$ (eV) & $\Delta E^{\mathrm{vert}}(r_\mathrm{HS})$ (eV) &  $\Delta E^{\mathrm{adia}}$ (eV)\\ \hline\hline
DMC ($\Delta \tau=0.0125$ a.u)& $0.65(3)$ &$-1.90(2)$ & $-0.36(4)$\\
DMC ($\Delta \tau=0.005$ a.u)& $0.65(3)$ & $-1.91(3)$ & $-0.36(4)$\\
\hline
\end{tabular}}
\caption{Adiabatic and vertical energy gaps for the [FeL$_2$]$^{2+}$ cation calculated with DMC at the single crystal experimental 
geometry. We report DMC results for two values of the imaginary time. The relative Monte Carlo statistical error is indicated in bracket. The 
experimental molecular structure used for the calculation is taken from the X-ray data of reference \cite{Money}.} 
\label{energy_SC_molecule_exp}
\end{table}
The argument proceeds then as follows. Firstly, one has to repeat the calculations using the experimental geometries measured for the
molecule in the crystal form~\cite{Money}. These are less symmetric and present shorter metal-ligand bond-lengths than those optimized in the 
vacuum (see Tab. \ref{Tab_bond_SC_molecule}). The result is that the DMC-calculated $\Delta E^{\mathrm{adia}}$ gets smaller, although
it maintains the negative sign (compare Tab.~\ref{energy_SC_molecule_exp} with Tab.~\ref{energy_SC_molecule_theory}). Secondly, the 
electrostatic potential felt by the molecule in the crystal and due to both the counter-ions and the other molecules needs to be taken into
account. This produces a relative shift of the HS and LS potential energy surfaces. The magnitude of this effect, which tends to stabilize 
the LS state, has been recently estimated \cite{Robert_2} to be of the order of $0.5$~eV. When the effect of the geometry and the electrostatic 
corrections are both included, DMC allows us to estimate a $\Delta E^{\mathrm{adia}}$ for the condensed phase of about 0.2~eV. This is now 
positive, i.e. the ground state is LS, and very close to the typical values of the adiabatic energy gap inferred from experimental data \cite{Reiher_3}.

We finally turn our attention to the assessment of the performaces of the various exchange-correlation functionals. Table~\ref{energy_SC_molecule_theory} 
displays the vertical and adiabatic energy gaps calculated with DFT. We note that BP86 underestimates the exchange so significantly that the molecule 
is predicted to be stable in the LS state ($\Delta E^{\mathrm{adia}}>0$). Furthermore the absolute value of 
$\Delta E^{\mathrm{vert}}(r_\mathrm{LS})$ [$\Delta E^{\mathrm{vert}}(r_\mathrm{HS})$] is much larger than the corresponding one computed with DMC. 
This means that the standard local density approximation predicts a very stable low spin ground state. B3LYP and PBE0 improve only slightly 
the accuracy of the calculated gaps and the LS state still remains massively over-stabilized. In contrast, as in the case of small Fe$^{2+}$ model 
complexes \cite{me2}, HH is found to be the functional that performs better, yielding a fair agreement with the DMC gaps. 

Importantly our analysis demonstrates that the assessment of the performances of a given DFT functional can be completely erroneous, if one
insists on comparing the total energies calculated for the gas phase directly to experiments. If this was done with our DFT data, we would have  
concluded, as other authors did \cite{Reiher_1,Reiher_2}, that B3LYP* was the best functional. Our analysis instead demonstrates that the correct assignment needs to be
done against a reliable benchmark, with the result that the best suited functional must carry a fraction of exact exchange near to 50\%.

In conclusion, we have shown that for a typical SC molecule in the gas-phase the ground state is high spin and discussed how the this 
becomes singlet in the condensed phase. We have also assessed the performaces of DFT against this problem, demonstrating that the 
HH hybrid functional, including $50\%$ of exact exchange, is able to provide a quite accurate estimate of the energetic of the molecule. 
Our results then finally shed light on the long-standing issue of establishing the ground state of SC complexes. 

{\sc acknowledgments:}
We thank N. Baadji for useful discussions. This work is sponsored by EU (Hints project). Computational resources have been provided by TCHPC and ICHEC.


\begin{thebibliography}{100}
\bibitem{note} SC molecules incorporating Fe$^{3+}$, Co$^{2+}$ and Mn$^{3+}$ have also been reported. However these are much less common then 
those containing Fe$^{2+}$ ions.
\end{thebibliography}
\end{document}